# On the benefits of the Eastern Pamirs for sub-mm astronomy

Alexander V. Lapinov*[a,b], Svetlana A. Lapinova[c,d], Leonid Yu. Petrov[e], Daniel Ferrusca[f]
[a]Institute of Applied Physics of the Russian Academy of Sciences, 46 Ulyanov Str., Nizhny Novgorod, Russia 603950; [b]Minin University, 1 Ulyanov Str., Nizhny Novgorod, Russia 603950; [c]Lobachevsky State University, 23 Gagarin Ave., Nizhny Novgorod, Russia 603950; [d]Higher School of Economics 25/12 Bolshaya Pecherskaya Str., Nizhny Novgorod, Russia 603155; [e]NASA GSFC, 8800 Greenbelt Rd., Greenbelt, MD 20771, USA; [f]Instituto Nacional de Astrofísica, Óptica y Electrónica, Luis Enrique Erro 1, Tonantzintla, Puebla, C.P. 72840, México

**ABSTRACT**

Thanks to the first mm studies on the territory of the former USSR in the early 1960s and succeeding sub-mm measurements in the 1970s – early 1980s at wavelengths up to 0.34 mm, a completely unique astroclimate was revealed in the Eastern Pamirs, only slightly inferior to the available conditions on the Chajnantor plateau in Chile and Mauna Kea. Due to its high plateau altitude (4300 – 4500 m) surrounded from all sides by big (~7000 m) air-drying icy mountains and remoteness from oceans this area has the lowest relative humidity in the former USSR and extremely high atmospheric stability. In particular, direct measurements of precipitated water vapor in the winter months showed typical pwv=0.8 – 0.9 mm with sometimes of 0.27 mm. To validate previous studies and to compare them with results for other similar regions we performed opacity calculations at mm – sub-mm wavelengths for different sites in the Eastern Pamirs, Tibet, Indian Himalayas, APEX, ALMA, JCM, LMT and many others. To do this we integrate radiative transfer equations using the output of NASA Global Modeling and Assimilation Office model GEOS-FPIT for more than 12 years. We confirm previous conclusions about exceptionally good astroclimate in the Eastern Pamirs. Due to its geographical location, small infrastructure and the absence of any interference in radio and optical bands, this makes the Eastern Pamirs the best place in the Eastern Hemisphere for both optical and sub-mm astronomy.

## INTRODUCTION

Nowadays, thanks to modern high-altitude observatories, amazing results of ground based sub-mm astronomy were obtained, the peak of which appears to be the measurement of the black hole horizon in the galaxy M87[1]. At the same time, it should be noted, that almost all EHT instruments are located in the Western Hemisphere. And this is despite the fact that there are places in the Eastern Hemisphere that are very suitable for similar studies. The above fact significantly limits not only the capabilities of the Event Horizon Telescope, but also many other sub-millimeter studies. In addition, it is well known that January and February are the wettest months for APEX and ALMA, when their observations are practically stopped. All this gives reason to think about the search for possible additional places in the northern part of the Eastern Hemisphere, comparable in terms of astroclimate. And there really are such places. Historically, thanks to the first mm studies on the territory of the former USSR in the early 1960s, a completely unique astroclimate was revealed in the Eastern Pamirs. The pioneers of such studies were researchers from the Nizhny Novgorod Radiophysical Institute, which carried out their measurements in different parts of the Eastern Tadjikistan from June to November 1962 with their equipment at 1.3 and 1.8 mm[2-3] in rather complicated conditions (see Fig.1). After some time, part of the staff of these studies founded the department of radio astronomy at the Institute of Applied Physics of the RAS, N.Novgorod.

Later, subsequent studies were performed in the 1970s - early 1980s in the Eastern Pamirs at the Shorbulak observatory (4350 m altitude) equipped with an optical 70 cm reflector at sub-mm wavelengths up to 0.34 mm and in optics[4-6]. These studies were accompanied by direct measurements of the water vapor content in the air by daily launches of meteo probes (up to 4 times per day).

*lapinov@ipfran.ru; phone +7 952 785-2211; fax +7 831 418 9040

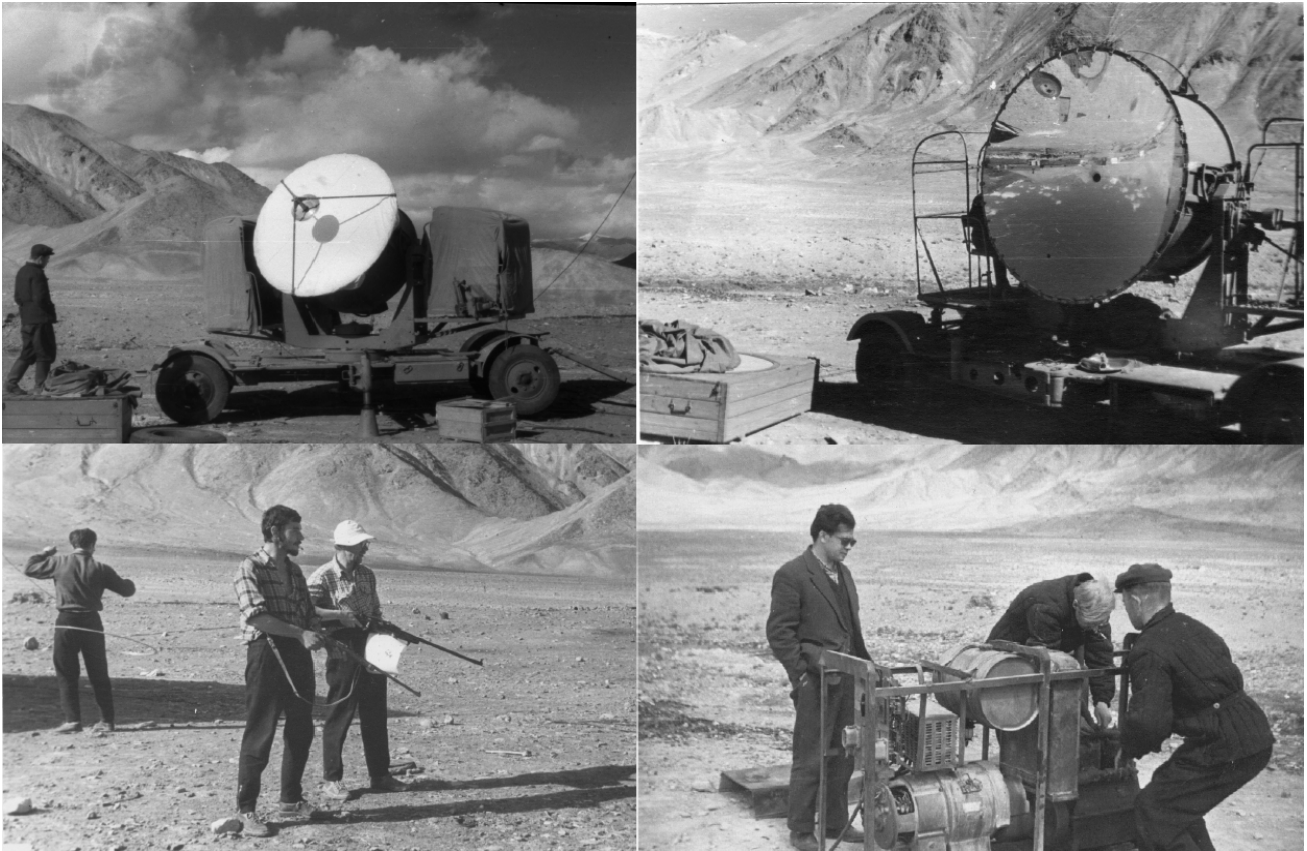
Figure 1. The first mm studies in the Eastern Pamirs during the expedition of the Nizhny Novgorod Radiophysical Institute (June-Nov. 1962). All the photos are kindly provided by L.V. Lubyako participated in these measurements.

The main conclusions of these studies (their small part) were as follows.
1. The Eastern Pamirs has the driest climate in the USSR (less than 100 mm of precipitation per year, there is absolutely no grass cover; in most cases – completely snowless winters); total absence of dust; the average annual temperature is -3°C, daily difference is less than 6°C; average wind ~ 6m/s, subsiding at night; over 100 completely clear nights per year.
2. Extinction in optics sometimes is better than on Mauna Kea, close to Rayleigh scattering.
3. An exceptionally favorable place for sub-mm studies; average winter pwv = 1.2 mm (in Hawaii – 1.9 mm), sometimes pwv = 0.27mm.

To validate previous conclusions and to compare them with results for other similar regions we performed opacity calculations at mm – sub-mm wavelengths for different sites in the Eastern Pamirs, Tibet, Indian Himalayas, APEX, ALMA, JCMT, LMT and many others. Starting with this article, we are planning a series of publications based on the results of our astroclimate analysis for a large number of both operating observatories and those that are planned for the near future. To perform our calculalions for the Eastern Pamirs we selected 3 sites with similar conditions and altitude: Koluch-Kul – 4475 m, Shorbulak – 4289 m (both in Tajikistan), Muztagh-Ata – 4526 m (China). In all cases, the given altitudes and selected geographical coordinates were obtained from high-resolution Google maps. Since three above sites are located at similar altitudes and astroclimate conditions, in this work we will focus mainly on Koluch-Kul. Some time ago, a small plateau of ~1 km$^2$ in this site at an altitude of 4260 m above sea level at the foot of a high gentle hill was selected by Lebedev Physical Institute of the RAS as the most promising place for a cosmic ray station. This was done due to the practically snowless winters and the presence of a spring with drinking water. And the top of the hill, thanks to the preliminary infrastuctre, may be considered as the most promising place for the sub-mm observatory. Fig. 2 shows the surroundings at Koluch-Kul and the Shorbulak observatory. Note that, in addition to the above, in the Eastern Pamirs there are a large number of similar places in the territory, which is more than an order of magnitude larger than the ALMA.

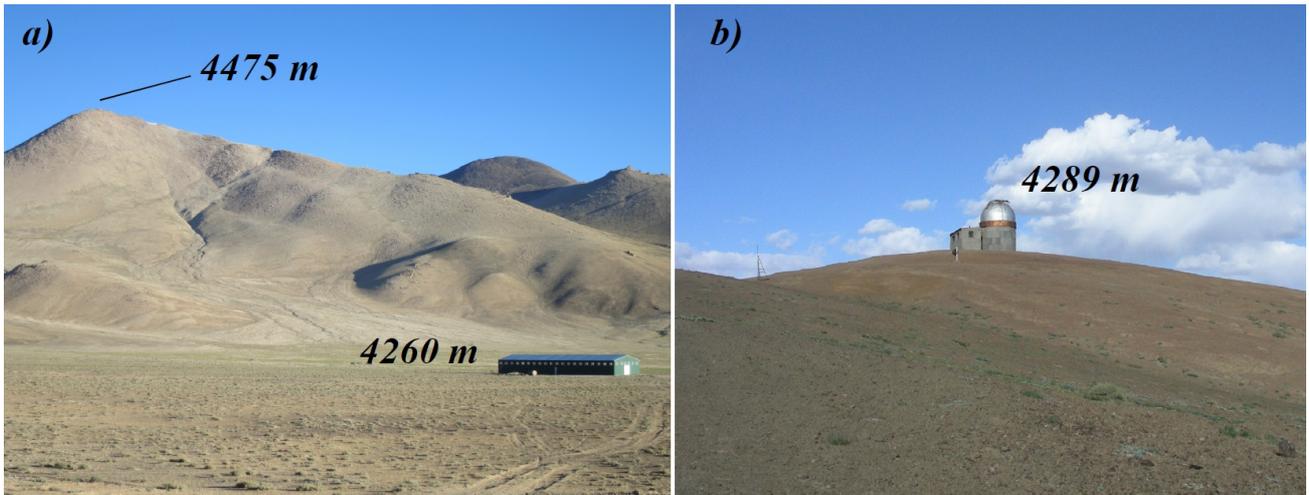

Figure 2. Surroundings nearby a) Koluch-Kul Cosmic Ray Station of the Lebedev Physical Institute of the RAS, b) Shorbulak observatory. Both photos are kindly provided by A.S. Borisov, a head of the Pamirs Cosmic Ray Station.

## METHODOLOGY

To calculate the atmosphere transmission for individual sites we use the approach described in detail previously for modeling path delay in the neutral atmosphere[7]. Since the mixing ratio of water vapor, which plays a determining role for sub-mm opacity in atmosphere, is highly variable, its distribution cannot be deduced from direct surface measurements. For this reason, we integrate radiative transfer equations by using output of the NASA Global Modeling and Assimilation Office model GEOS-FPIT[8]. Corresponding data are now publicly available, and really, at the moment, there are several centers in the world that produce similar outputs. They evaluate atmosphere parameters using various ground, air-born, and space-born measurements that are assimilated into a dynamic model. The output of this model defines the parameters of the state of the atmosphere on a 4-dimensional grid. Three parameters are important for reduction of astronomy observations: air temperature $T$, total atmospheric pressure $P$, and partial pressure of water vapor P$w$. Current models have 72 levels in height, 0.25°× 0.3125° spatial greed and 3 hours resolution in time. After that the absorption through the atmosphere at any selected frequency can be calculated for any selected place by using standard spectroscopic parameters listed in recommended databases[9]. To perform our analysis we calculated atmosphere opacity for more than 40 different sites at frequencies from about 18 GHz to 345 GHz. Some places were selected to make a comparison with the measurements of atmosphere opacity by a chopper-wheel method with radio telescopes that we have used since 2008. Sometimes, this comparison is possible due to available tau-meters at the observatories.

In particular, Fig. 3 shows a comparison of 225 GHz zenith opacity at LMT-50m directly measured in 2019 January by a tip radiometer (see details, for example, in a set of previous articles[10,11]) and calculated by us from the output of NASA global weather numerical model GEOS-FPIT. Green spikes are due to clouds at sky dips. This figure shows only a small portion of the data we used for comparison. Currently LMT is not a supplier of weather parameters to the global meteorological network. In spite of that, Fig. 3 demonstrates a fairly good agreement between model calculations and direct measurements. Additionally we obtained a very good coincidence when comparing chopper-wheel measurements at similar frequency for the IRAM-30m and at 86.2 GHz for the 20m Onsala telescope. We plan to discuss these results in more detail in future papers. The general conclusion from the comparison of our calculations and measurements is that, within the errors, their results are coincident. Of course, at the moment, global weather models cannot predict the detailed cloud structure in a chosen direction in the sky. For this reason, cloud absorption is not taken into account in our calculations. On the other hand, when analyzing data from tau-meters, all spikes are filtered out. So, it is possible to make a bold statement that nowadays, in order to judge in advance about the atmosphere transmission in any particular site at a chosen frequency, for a preliminary assessment it is not necessary to install special equipment and make detailed measurements. This can be done with a sufficiently high accuracy by our method.

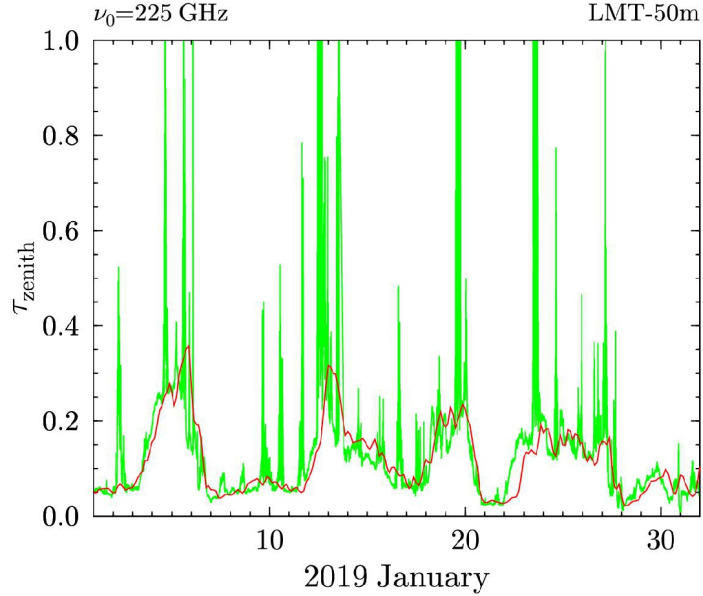

Figure 3. 225 GHz zenith opacity at LMT-50m directly measured by a tip radiometer (green) and calculated from the output of NASA global weather numerical model GEOS-FPIT. (red). Green spikes are due to clouds at sky dips.

## RESULTS AND DISCUSSION

As it was stated in previous studies[4-6], the quality of the astroclimate in the Eastern Pamirs meets all the best requirements for ground-based sub-mm measurements. To validate this conclusion we performed comparisons of the atmospheric transparency for Koluch-Kul, Shorbulak and Muztagh-Ata with such observatories as APEX, LMT-50m, JCMT-15m, IRAM-30m, different sites proposed for the sub-mm observatories in Tibet, Indian Himalayas, Suffa and a number of other places. Despite the fact that our calculations were performed for all observatories for a wide frequency range from 18 to 345 GHz, in this work we will focus mainly on the analysis at 230 GHz. This frequency corresponds to the CO J=2–1, which is one of the most popular transition in large-scale studies of star-forming regions, and is the frequency that was used by the EHT for the M87 black hole horizon measurements. Meanwhile, it is possible to note that in order to find the optical depth of the atmosphere at 129 GHz, corresponding to the SiO J=3–2 near the minimum of 2 mm transparency window, we just need to divide $\tau_{230\ GHz}$ by 3. And vice versa, in order to find the optical depth of the atmosphere at a frequency of 345 GHz, corresponding to the CO J=3–2 near the center of the 0.8 mm window, we just need to multiply $\tau_{230\ GHz}$ by 3.

Since the astroclimate may vary from year to year, we performed our analysis for each of the site for the period from 01.01.2008 to 31.12.2019. In addition, we have made our analysis for each month separately and with averaging each month over 12 years. Because our calculations were carried out with a step of 3 hours, the statistics are representative and allow us to plot probability distributions. As a result, in Fig. 4, we show monthly distributions of the probability density for zenith opacity and their cumulative distributions at 1.3 mm for Koluch-Kul on the Eastern Pamirs (blue and green, correspondingly) in comparison with APEX (black and red) averaged over 12 years. It is quite clear that starting from April up to November inclusively, APEX is out-of-competition. At the same time, all January and February, a clear win belongs to Koluch-Kul. Stably good conditions in the Eastern Pamirs, when zenith transparency at 230 GHz in most cases is higher than 90%, are from November to March. The best time is December - February. This is exactly the period when APEX and ALMA practically stop working. Thus, both sites complement each other quite well. Figs. 5 and 6 show that the conditions in the Eastern Pamirs are always better than those available at LMT and Pico Veleta. Among other things, due to the protection from all sides by high ~7000 m mountains, there are here no such strong storms as at LMT.

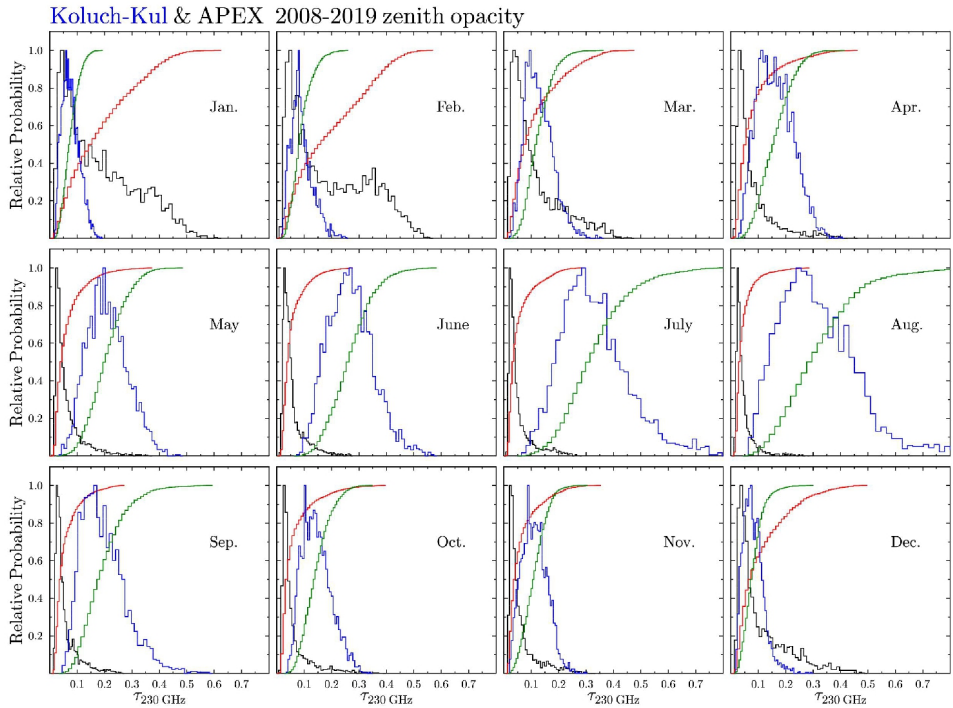

Figure 4. Probability density and cumulative distributions of 230 GHz zenith opacity at Koluch-Kul on the Eastern Pamirs (blue and green, correspondingly) in comparison with APEX (black and red) calculated from the output of NASA global numerical weather model GEOS-FPIT for 12 years (01.01.2008-31.12.2019).

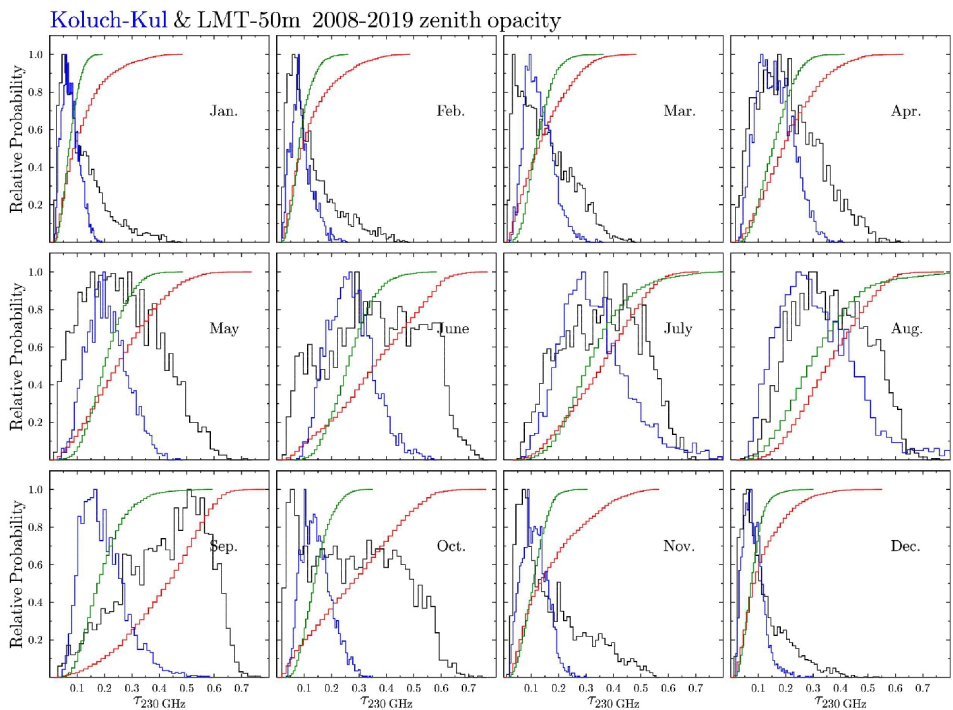

Figure 5. Probability density and cumulative distributions of 230 GHz zenith opacity at Koluch-Kul on the Eastern Pamirs (blue and green, correspondingly) in comparison with LMT-50m (black and red) calculated from the output of NASA global numerical weather model GEOS-FPIT for 12 years (01.01.2008-31.12.2019).

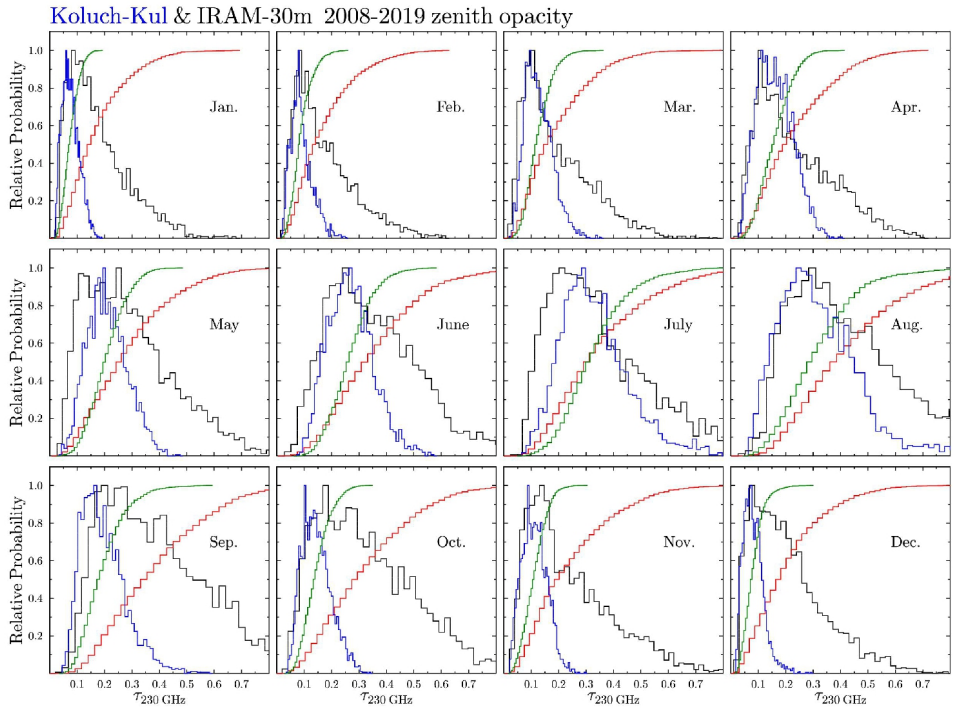

Figure 6. Probability density and cumulative distributions of 230 GHz zenith opacity at Koluch-Kul on the Eastern Pamirs (blue and green, correspondingly) in comparison with IRAM-30m (black and red) calculated from the output of NASA global numerical weather model GEOS-FPIT for 12 years (01.01.2008-31.12.2019).

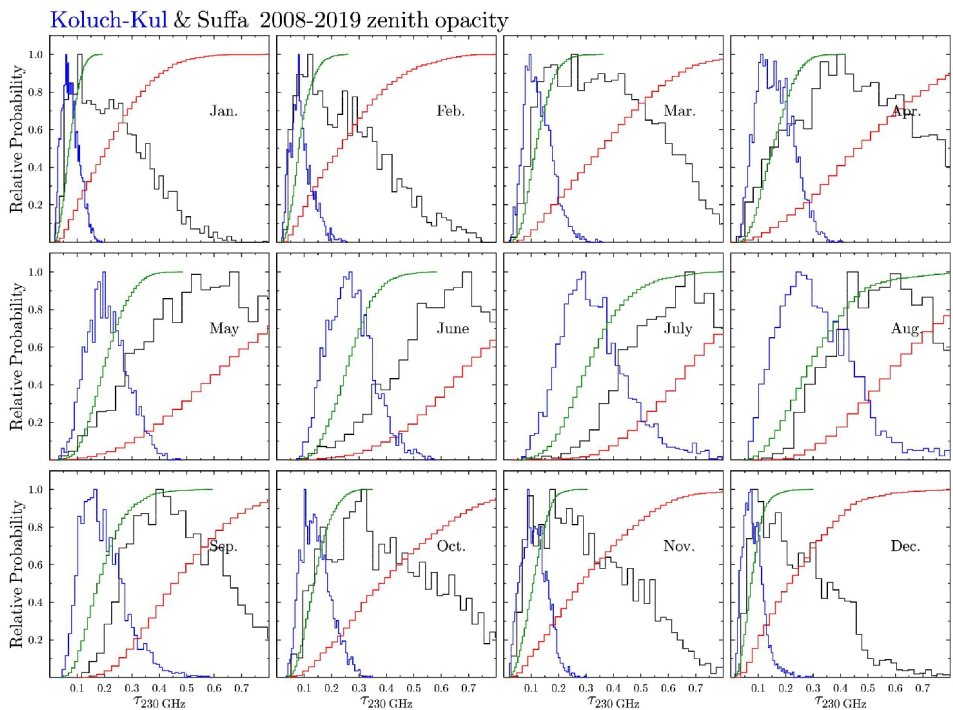

Figure 7. Probability density and cumulative distributions of 230 GHz zenith opacity at Koluch-Kul on the Eastern Pamirs (blue and green, correspondingly) in comparison with Suffa (black and red) calculated from the output of NASA global numerical weather model GEOS-FPIT for 12 years (01.01.2008-31.12.2019).

Finally, Fig. 7 shows a comparison of the optical thickness at the zenith in Koluch-Kul and in the Suffa plateau. The latter site was chosen over 40 years ago and was later considered by the astronomical leadership of Russia as an alternative to ALMA for the measurement up to 0.8 mm. It is located in Uzbekistan at an altitude of 2300m. This is almost the same as for the SEST-15m, located in a more driest climate and closed in 2003. It can be seen that the transparency of the atmosphere on the Suffa plateau cannot be compared with the Eastern Pamirs at all and is not suitable for good observations even at 230 GHz. An additional bad thing for Suffa and the nearby Maidanak (~2600 m altitude) is the heavy fractional cloud cover from November to May[12], in contrast to the atmosphere transparency without clouds. Nonetheless, nowadays the ground based radio astronomy is in great need of very large diameter instruments at 3 mm wavelength. For this reason, taking into account the already invested funds, an antenna with a diameter of 70 m for 3 and possibly 2 mm would be in great demand.

As an example of seasonal behavior, in Fig. 8 we show the results of optical depth averaging over the last 12 years for four different observatories. Red, black and blue are the values for the first quartile, median and the third quartile boundaries. From top to bottom, the zenith optical depths of the atmosphere are shown at frequencies of 129, 230, and 345 GHz, respectively. As noted earlier, it is seen that to get the optical depth of the atmosphere at 129 GHz, we just need to divide $\tau_{230\,GHz}$ by 3, and in order to find the optical depth at 345 GHz, we just need to multiply $\tau_{230\,GHz}$ by 3. What about Ali sub-mm observatory in Tibet with ~500m higher altitude, in the winter season their conditions are better than in Koluch Kul and are very similar to best conditions in ALMA. At the same time, due to high summer humidity, their conditions (and especially atmosphere transparency in the Hanle, India) are much worse than in the Eastern Pamirs.

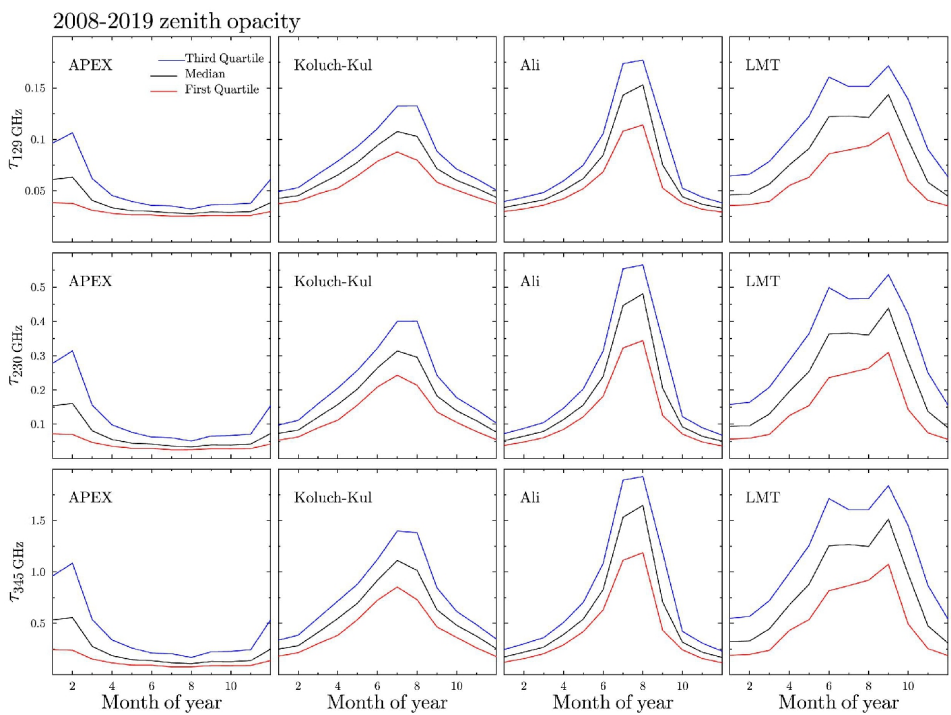

Figure 8. Seasonal variation in the zenith opacity for APEX (Chile), Koluch-Kul (Eastern Pamirs), Ali (Tibet) and LMT (Sierra Negra) at 129 GHz (top), 230 GHz (center) and 345 GHz (bottom) calculated from the output of the NASA GEOS-FPIT for 12 years period (01.01.2008-31.12.2019).

As for the annual variations, the astroclimate of the Eastern Pamirs is extremely stable. This is in stark contrast, for example, with the atmosphere transparency in Hawaii, where conditions can vary greatly from year to year. In particular, the JCTM opacity was exceptionally good throughout most of 2010, corresponding to the best conditions for APEX in August. At the same time, throughout 2018, the atmosphere transparency was almost an order of magnitude worse.

All these features in relation to individual telescopes and detailed statistics for different wavelengths, taking into account the specifics of each instrument, we plan to discuss in the forthcoming papers.

# CONCLUSIONS

Currently, there are a number of first-class mm and sub-mm observatories in the Western Hemisphere, which have proven themselves to be excellent in observations at ~900 GHz[13,14] and even at 1.5 THz[15]. Some of them were very successful as parts of the EHT project. There are also a number of projects to launch mm radio telescopes into space. Unfortunately, the latter, despite the smaller antennas and worse angular resolution in comparison with ground-based instruments, are several orders of magnitude more expensive, their development and manufacturing is much more time consuming, a successful launch is not guaranteed, and in a case of any breakdown, its elimination is almost impossible. For a long time, choosing a good site for new observatories has been a complex and resource-consuming task. The mistakes made in this case were very expensive. With this work, we have shown that nowadays the last problem concerning the analysis of the astroclimate can be solved at a much lower cost. We have shown clearly that the Eastern Pamirs and Tibet in the Eastern Hemisphere are indeed a good supplementary sites to the Chajnantor plateau and Mauna Kea. As a first step , we call on the astronomical community to pay attention to Koluch-Kul, Shorbulak and Mustagh-Ata, which may become the first points of the future submillimeter antenna array with an angular resolution that is an order of magnitude higher than ALMA.

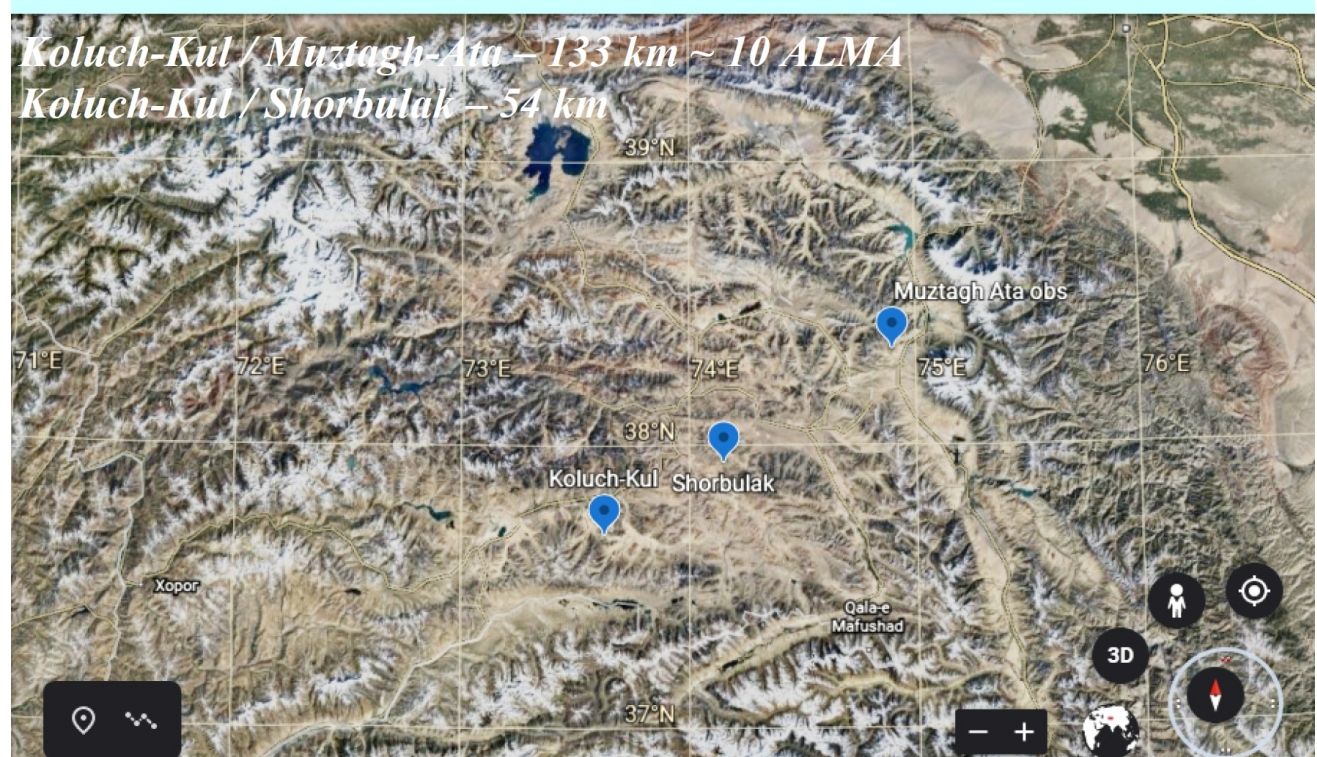

Figure 9. First proposed sub-mm sites for Pamirs Large MM Array, PaLMA. Google.ru site view of the Eastern Pamirs plateau surrounded by mountain ranges of ~7000 m.

# ASKNOWLEGMENTS

It is our pleasure to thank all people, who advised us on conditions in the Eastern Pamirs and provided us with corresponding photos. The research of AL was carried out within the IAP RAS state program 0035-2019-0005.

# REFERENCES

[1] EHT Collaboration et al., "First M87 Event Horizon Telescope Results. I. The Shadow of the Supermassive Black Hole," ApJL, 875, L1 (2019).
[2] Fedoseev, L. I., "Radioemission of the Moon and Sun at 1.3 mm wavelength," Radiophysics and Quantum Electronics, 6, 655-659 (in Russian) (1963).
[3] Naumov, A. I., "Radioemission of the Moon and Sun at 1.8 mm wavelength," Radiophysics and Quantum Electronics, 6, 848-849 (in Russian) (1963).
[4] Sholomitskii, G. B., Maslov, I. A., Grozdilov, V. M., "Submillimeter transmission of the atmosphere at Shorbulak, Eastern Pamirs," Sov. Astron., 26(3), 358-361 (1982).
[5] Kanaev, I. I., Sholomitskii, G. B., Maslov, I. A., Grozdilov, V. M., "Shorbulak in the Eastern Pamirs - a Promising Site for Astronomical Observations," ASPRv, 3, 329 (1984).
[6] Maslov, I. A., Soglasnova, V. A., Sholomitskii, G. B., Gromov, V. D., Nikolskii, Y. V., Maslennikov, K. L., "Submillimeter Spectrophotometry in the Pamirs," SvAL, 15, 287-290 (1989).
[7] Petrov, L., "Modeling of path delay in the neutral atmosphere: a paradigm shift," 2015arXiv150206678P (2015).
[8] NASA Global Modeling and Assimilation Office, http://gmao.gsfc.nasa.gov.
[9] The International Telecommunication Union Recommendation P.676, "Attenuation by atmospheric gases and related effects," https://www.itu.int/rec/R-REC-P.676/en
[10] Ferrusca, D. and Contreras R., J., "Weather monitor station and 225 GHz radiometer system installed at Sierra Negra: the Large Millimeter Telescope site," Proc. SPIE 9147, 914730 (2014).
[11] Zeballos, M., Ferrusca, D., Contreras R., J., Hughes, D. "Reporting the first 3 years of 225 GHz opacity measurements at the site of the Large Millimeter Telescope Alfonso Serrano," Proc. SPIE 9906, 99064U-1 (2016).
[12] Hellemeier, J. A., Yang, R., Sarazin, M., Hickson, P., "Weather at selected astronomical sites – an overview of five atmospheric parameters," MNRAS, 482, 4941–4950 (2019).
[13] Schilke, P., Mehringer, D. M., and Menten, K. M., "A submillimeter HCN laser in IRC +10216," ApJ, 528, L37-L40 (2000).
[14] Schilke, P., and Menten, K. M., "Detection of a second, strong submillimeter HCN laser line toward carbon stars," ApJ, 583, 446-450 (2003).
[15] Wiedner, M. C., Wieching, G., Bielau F., et al., "First observations with CONDOR, a 1.5 THz heterodyne receiver," A&A, 454, L33–L36 (2006).